\documentclass[a4paper]{iopart}
\usepackage{graphicx}
\usepackage{epstopdf}
\usepackage{subfigure}
\usepackage{hyperref}
\usepackage{caption}
\usepackage{tabularx}
\usepackage{layouts}
\usepackage{dcolumn}
\usepackage{bm}
\usepackage{amsmath}
\usepackage{dsfont}
\usepackage{comment}
\usepackage{bbold}
\DeclareMathOperator{\di}{d\!}
\usepackage{cite}
\usepackage[margin=1in]{geometry}
\graphicspath{ {./Figures/} }

\newcommand{\bmath}{\begin{displaymath}}
\newcommand{\emath}{\end{displaymath}}

\newcommand{\be}{\begin{equation}}
\newcommand{\ee}{\end{equation}}
\newcommand{\bea}{\begin{eqnarray}}
\newcommand{\eea}{\end{eqnarray}}

\newcommand{\bmultl}{\begin{multline}}
\newcommand{\emultl}{\end{multline}}

\newcommand{\bsubeq}{\begin{subequations}}
\newcommand{\esubeq}{\end{subequations}}
\newcommand{\bitemize}{\begin{itemize}}
\newcommand{\eitemize}{\end{itemize}}
\newcommand{\ket}[1]{\left|{#1}\right\rangle}
\newcommand{\bra}[1]{\left\langle{#1}\right|}
\newcommand{\abs}[1]{\left|{#1}\right|}

\newcommand{\bmx}{\begin{bmatrix}}
\newcommand{\emx}{\end{bmatrix}}
\newcommand{\bsmx}{\begin{smallmatrix}}
\newcommand{\esmx}{\end{smallmatrix}}

\begin{document}
\title{Probing bath-induced entanglement in a qubit pair by
measuring photon correlations}


\author{Ovidiu Cotlet$^1$ and Brendon W. Lovett$^1$}

\address{$^1$ School of Physics and Astronomy, University of St Andrews, KY16 9SS, UK}
\
\ead{ocotlet@gmail.com}
\ead{bwl4@st-andrews.ac.uk}

\begin{abstract}
Self-assembled quantum dots are ideal structures in which to test theories of open quantum systems: Confined exciton states can be coherently manipulated and their decoherence properties are dominated by interactions with acoustic phonons. We here describe the interaction of a pair of un-coupled, driven, quantum dot excitons with a common phonon environment, and find that this coupling effectively generates two kinds of interaction between the two quantum dots: An elastic coupling mediated by virtual phonons and an inelastic coupling mediated by real phonons. We show that both of these interactions produce steady state entanglement between the two quantum dot excitons. We also show that photon correlations in the emission of the quantum dots can provide a signature of the common environment. Experiments to demonstrate our predictions are feasible with the state-of-the-art technology and would provide valuable insight into quantum dot carrier-phonon dynamics. 
\end{abstract}

\section{Introduction}

Self-assembled quantum dots (QDs), quantum heterostructures in which electrons and holes are confined in all three dimensions, are artificial solid-state atoms with tailored optical and electronic coherence properties.
Impressive progress on fabrication and optical manipulation techniques has enabled high fidelity preparation, control and readout of the quantum states of charge carriers confined in individual QDs 
\cite{degreve13,godden12,kiravittaya2009advanced,berezovsky2008picosecond,ramsay2008fast,vamivakas2009spin,press2008complete,gerardot2008optical} and in QD pairs \cite{greilich11,gerardot2005photon,robledo2008conditional}. Indeed, entanglement of photons and carrier spins has been fully characterised~\cite{degreve13ncomm}, and quantum dots are now recognised as superb single and entangled photon sources~\cite{salter10,he13}. At the same time, the coupling of QDs to their solid-state environment~\cite{ramsay2010damping,ramsay2010phonon,kuhlmann13}
provides a rich platform for the study of open system effects that may be more difficult to observe in, for instance, atomic systems. 

There is considerable evidence that the decoherence effects induced by the solid state environment of QD excitons are dominated by interactions with longitudinal acoustic phonons via the deformation potential coupling \cite{ramsay2010damping,ramsay2010phonon}. The QD interaction with the phonon bath leads to pure dephasing of bare excitons, or relaxation of driven excitons, in individual QDs \cite{borri2005exciton,krummheuer2002theory,jacak2003coherent} -- but a range of not-yet-observed phenomena are predicted to appear due to the interaction of multiple QDs with a phonon bath. For example, phonon induced dephasing results in entanglement decay between two QDs at a much faster rate than the individual QD dephasing rate \cite{yu04,dodd2004disentanglement}. Moreover, there has been considerable interest in phonon-assisted processes that appear when coupled QDs interact with the same phonon bath. Phonon assisted tunneling \cite{heitz1998excitation,tackeuchi2000dynamics,nakaoka2006direct} , relaxation \cite{grodecka2008phonon,wijesundara2011tunable} and excitation transfer \cite{gerardot2005photon,ortner2005energy,nishibayashi2008observation} have been demonstrated. However, in this paper we focus on an investigation of the phenomena that appear solely due to the collective interaction of {\it un}-coupled QDs with a common phonon reservoir, which cannot be explained by an interaction with separate reservoirs. 

The properties of the phonon bath are often investigated through excitonic occupation dynamics \cite{mogilevtsev2008driving,forstner2003phonon,vagov2007nonmonotonic,machnikowski2004resonant}. Here we will study another powerful method -- the analysis of emitted photon statistics. Owing to the strong optical transition dipole of semiconductor QDs, the optical properties of individual QDs \cite{xu2007coherent,ates2009post,flagg2009resonantly,vamivakas2009spin,ulhaq2012cascaded}, as well as ensembles of QDs  \cite{flagg2011interference,patel2010two,scheibner2007superradiance} have been studied extensively with this technique. For example, the
single qubit second-order fluorescence intensity correlation function $g^{(2)}$ has been investigated both experimentally \cite{muller2007resonance} and theoretically \cite{nazir2008photon} and has been shown to yield important information about the nature of the QD solid-state environment. Furthermore, the two QD intensity correlation function has been used to characterize the coupling mechanism between two adjacent dots \cite{gerardot2005photon,stinaff2006optical,beirne2006quantum}. 

In this paper we will show that photon statistics measurements can be used to probe the QDs immediate environment and find signatures of a common environment  in a pair of un-coupled, driven QDs. We will describe how the interaction with a common phonon field results in both a coherent elastic coupling mediated by virtual phonons and an inelastic incoherent coupling with emission and absorption of real phonons. Although the interaction of excitons with phonons usually results in decoherence and entanglement decay we find that when driven dots interact with the same environment this interaction can be used to entangle the QDs, even in the steady state. This complements previous work in which environment-induced entanglement of undriven, and {\it unseparated} atoms was seen to persist {\it ad infinitum}~\cite{benatti05}; for our QD case the assumption of zero separation is obviously unrealistic. A further study of 
{\it undriven}, spatially separated dots found environment-induced entanglement decays away to zero, albeit on very long timescales~\cite{mccutcheon09}.
We find that intensity correlation measurements of emitted photons provide a signature of the common environment and can be used to measure the strength of the coherent and incoherent coupling mechanisms.


In the following section we present the Hamiltonian of the QDs coupled to the radiation and solid-state environment. In Section \ref{sec:Method} we trace out the radiation and phonon bath in a Born-Markov fashion to obtain a second-order master equation in Lindblad form for the reduced density matrix of the two QDs.  We go on in Section \ref{sec:Steady} to investigate the steady state solution of the master equation showing that the common phonon bath induces entanglement, before in Section~\ref{sec:Photon} finding signatures of a common environment in the normalized intensity correlation function $g^{(2)}$.
We conclude in Section \ref{sec:Conclusion}.

\section{Model}\label{sec:Model}
The system we consider is formed of two driven QDs that do not interact with each other directly, but which are coupled to the same phonon and radiation bath.  We will consistently use the parameters of a GaAs self-assembled exciton QD, although all the calculations can apply to any optical emitters in a solid state environment, so long as the approximations we used are valid. In this section we obtain the Hamiltonian for our system following closely the derivation in \cite{kok2010introduction}. 

We model a single QD  as a two level system with ground state $\ket{0}$ and excited state $\ket{\xi}$ separated by an energy difference $\omega_{\xi}$. We define the creation operator for an exciton as $c^\dagger = \ket{\xi}\bra{0}$ with the annihilation operator its Hermitian conjugate. We will denote an operator $O$ acting on the Hilbert space of the first dot (second dot) as $O_A \equiv  O \otimes I$ ($O_B \equiv I \otimes O$) or, when notational clarity demands, as $O^A \equiv  O \otimes I$ ($O^B \equiv I \otimes O$). 
Each dot is driven by its own near-resonant laser of frequency $\omega_l^j$which couples to each dot with a strength leading to a Rabi frequency $\Omega^j$ with $j\in\{A,B\}$. 

The dots are also coupled to the electromagnetic environment, which we represent as a bath of harmonic oscillators of frequencies $\Theta_{\bm{q}}$ and creation operators $a^\dagger_{\bm{q}}$, where $\bm{q}$ denotes a mode of the radiation field with wave vector ${\bm q}$. The coupling Hamiltonian is: 
\bea
H_I=\sum_{j\in\{A,B\}} (c_j^\dagger + c_j) \sum_{ \bm{q}}  f_{\bm{q}}^j (a_{\bm{q}}^\dagger + a_{\bm{q}}).
\eea
The interaction between the dots and the radiation field is fully characterized by the photon spectral density function: 
\bea
J_{\gamma}^j (\omega) \equiv 2 \pi \sum_{\bm{q}} \abs{f_{\bm{q}}^j}^2 \delta(\omega - \omega_{\bm{q}}).
\eea
For our calculation the spectral density will not vary significantly across the relevant frequencies and so we can consider it a constant,  $J_\gamma^j(\omega) \approx 1 / T^*$, with $T^*~\approx$~1~ns.  

QDs exist in a solid state environment, and so are also coupled to a common phonon bath. This is again represented as a collection of harmonic oscillators of frequencies $\omega_{\bm{k}}$ and creation operators $b^\dagger_{\bm{k}}$ where $\bm{k}$ denotes the phonon wave vector. By assuming the strong confinement limit, we may model the electron-hole wave functions $\psi_{e,h}$ as (unentangled) product states of the corresponding single particle wave functions. The essential physics of QDs may then be captured by assuming Gaussian single particle spatial wave functions
with standard deviation lengths $d_e$ and $d_h$ for electrons and holes respectively. The interaction of such states with longitudinal acoustic phonons coupled through the deformation potential is dominant\cite{ramsay2010damping}, and thus we obtain the following exciton-phonon interaction Hamiltonian: 
\bea
H_I=\sum_{j\in\{A,B\}} c_j^\dagger c_j \sum_{\bm{k}} t_{\bm{k}}^j ( b_{\bm{k}}^\dagger + b_{\bm{k}}),
\eea
where 
\bea
t_{\bm{k}}^A&=&i \sqrt{\frac{1}{2 \mu V \omega_{\bm{k}}}} \abs{\bm{k}} \left[ {\mathsf D}_e \exp\left(-\frac{d_e^2 \abs{\bm{k}}^2}{4}\right) -{\mathsf D}_h \exp\left(-\frac{d_h^2 \abs{\bm{k}}^2}{4}\right)  \right]
\eea
and 
\bea
t_{\bm{k}}^B&=&t_{\bm{k}}^A e^{i \bm{k} \cdot \bm{d}},
\eea
where $\mu$ is the GaAs mass density, $V$ is the volume of the crystal, ${\mathsf D}_{e,h}$ are the electron/hole deformation potential constants. We are also able to define a phonon spectral density: 
\bea
J_p (\omega) \equiv 2\pi  \sum_{\bm{k}} \abs{t_{\bm{k}}^j}^2 \delta (\omega - \omega_{\bm{k}} )  = \Lambda_{\mathds D} \left( \frac{\omega}{\omega_0}\right)^{\mathds D} {\cal P }^2 (\omega),
\eea
where $\omega_0$ is a scaling energy introduced for convenience and its value defined  in Table~\ref{tab:parameters}, along with all the other parameters we have used. ${\mathds D}=1,2,3$ is the dimensionality of the phonons and ${\cal P}^2(\omega)$ is the form factor given by: 
\bea
{\cal P}^2 (\omega) = \left(\frac{1}{{\mathsf D}_e-{\mathsf D}_h}\right)^2  \left({\mathsf D}_h^2 e^{-\omega^2/\omega_h^2} +{\mathsf D}_e^2 e^{-\omega^2/ \omega_e^2} - 2 {\mathsf D}_h {\mathsf D}_e e^{-\omega^2/\omega_{eh}^2} \right),
\eea
where $\omega_\alpha= c_s \sqrt{2}/ d_\alpha$ for $\alpha\in\{e,h\}$ and $\omega_{eh}=2 c_s / \sqrt{d_e^2 + d_h^2}$. The dimensionality dependent constant $\Lambda_{\mathds D}$ is given by: 
\bea
\frac{\Lambda_{\mathds D}}{\omega_0^{\mathds D}} = 
\begin{cases}
({\mathsf D}_e-{\mathsf D}_h)^2 / (2 \pi \mu_3 c_s^5), & \mathrm{when }\quad {\mathds D}=3,\\ 
({\mathsf D}_e-{\mathsf D}_h)^2 / (2  \mu_2 c_s^4), & \mathrm{when }\quad {\mathds D}=2,\\ 
({\mathsf D}_e-{\mathsf D}_h)^2 / (\mu_1 c_s^3), & \mathrm{when }\quad {\mathds D}=1.\\ 
\end{cases}
\eea
$\mu_{1, 2, 3}$ are the mass densities for different dimensionalities, and $c_s$ is the speed of sound.

The resulting Hamiltonian takes the form ($\hbar=1$):
\bea\label{StartingHamiltonian}
H&=&\nonumber \sum_{j\in \{A,B\}} \left[ \omega_\xi^j c_j^\dagger c_j+ 
\frac{1}{2}\Omega^j \cos(\omega_l^j t) (c^\dagger_j + c_j) \right]+\sum_{\bm{q}} \Theta_{\bm{q}} a^\dagger_{\bm{q}}a_{\bm{q}}+\sum_{\bm{k}} \omega_{\bm{k}} b^\dagger_{\bm{k}}b_{\bm{k}}\\
& &+\sum_{j\in \{A,B\}} (c^\dagger_j +c_j)  \sum_{\bm{q}} g^j_{\bm{q}} (a^\dagger_{\bm{q}} +a_{\bm{q}})+\sum_{j\in \{A,B\}} c^\dagger_j c_j  \sum_{\bm{k}} t^j_{\bm{k}} (b^\dagger_{\bm{k}} +b_{\bm{k}}).
\eea

In order to eliminate the time dependence in the QD part of the Hamiltonian we move to a rotating frame by applying the unitary transform $U=\exp(i H_0 t)$ with $H_0= \sum_{j\in \{A,B\}} \omega_l^j c^\dagger_j c_j$. After performing a rotating-wave approximation on the driving term and on the QD-radiative bath interaction,  the transformed Hamiltonian is: 
\bea
\tilde{H}&=&\nonumber \sum_j \left[ \Delta_e^j \tilde{c}_j^\dagger \tilde{c}_j+ \frac{1}{2}\Omega^j(\tilde{c}^\dagger_j + \tilde{c}_j) \right]+\sum_{\bm{q}} \Theta_{\bm{q}} a^\dagger_{\bm{q}}a_{\bm{q}}+\sum_{\bm{k}} \omega_{\bm{k}} b^\dagger_{\bm{k}}b_{\bm{k}}\\
& &+\sum_j  \sum_{\bm{q}} g^j_{\bm{q}} (\tilde{c}_j e^{-i\omega_l^j}a^\dagger_{\bm{q}} +\tilde{c}^\dagger_j e^{i \omega_l^j t}a_{\bm{q}})+\sum_j \tilde{c}^\dagger_j \tilde{c}_j  \sum_{\bm{k}} t^j_{\bm{k}} (b^\dagger_{\bm{k}} +b_{\bm{k}})
\eea
where we have denoted the interaction picture creation operators with $\tilde{c}_j^\dagger = e^{-i \omega_l^j} c_j^\dagger$.  and $\Delta_e^j = \omega_\xi^j-\omega_l^j$. We have now dropped the explicit $\{A, B\}$ indices over which $j$ is summed, but this will henceforth always be assumed.

Since the two QDs do not interact with each other directly we can diagonalize the DQD part of the Hamiltonian by diagonalizing each QD part separately. For each QD the resulting eigenenergies are $W_j=\sqrt{(\Delta_e^j)^2 +( \Omega^j)^2 }$; the resulting eigenstates are $\ket{e}_j= \cos(\theta_j/2) \ket{\xi}_j + \sin(\theta_j/2) \ket{0}_j$ and $\ket{g}_j=-\sin(\theta_j/2) \ket{\xi}_j + \cos(\theta_j/2) \ket{0}_j$, and we have introduced the mixing angles $\theta_j =\arccos(\Delta_e^j/ W_j)$. Using the Pauli spin notation ($\sigma_+ = \ket{e}\bra{g}$,  $\sigma_z = \ket{e}\bra{e} - \ket{g} \bra{g}$, $I = \ket{e}\bra{e} + \ket{g} \bra{g}$), we may now write the Hamiltonian as: 
\bea \label{Hamiltonian}
\tilde{H}&=&\nonumber \sum_j \frac{W_j}{2}\sigma_z^j +\sum_{\bm{q}} \Theta_{\bm{q}} a^\dagger_{\bm{q}}a_{\bm{q}}+\sum_{\bm{k}} \omega_{\bm{k}} b^\dagger_{\bm{k}}b_{\bm{k}}\\
& &\nonumber+\sum_j  \sum_{\bm{q}} g^j_{\bm{q}} \left\{ \left[ \sigma_+^j\cos^2 \frac{\theta_j}{2}  -\sigma_-^j \sin^2 \frac{\theta_j}{2} +\frac{\sin\theta_j}{2}\sigma_z^j \right] e^{i \omega_l^j t}a_{\bm{q}} +h.c. \right\}\\
& & +\sum_j \left\{- \frac{\sin\theta_j}{2} (\sigma_-^j + \sigma_+^j) + \frac{\cos\theta_j}{2}\sigma_z^j+\frac{1}{2} I ^j   \right\}\sum_{\bm{k}} t^j_{\bm{k}} (b^\dagger_{\bm{k}} +b_{\bm{k}}).
\eea
\bigskip
\begin{minipage}{\textwidth}
\centering
\bigskip
\captionof{table}{Properties for GaAs quantum dots }  \label{tab:parameters}
\begin{tabularx}{0.9\textwidth}{c  X }
\hline
Electron deformation potential ${\mathsf D}_e$ & $14.6$ eV\\
Hole deformation potential ${\mathsf D}_h$ & $4.8$ eV\\
Mass density in 3D $\mu_{3}$ & $5.0\cdot 10^3$ kg m$^{-3}$ \\ 
Mass density in 2D $\mu_{2}$ & $3.0\cdot 10^{-6}$ kg m$^{-2}$ \\ 
Mass density in 1D $\mu_{1}$ & $1.7\cdot 10^{-15}$ kg m$^{-1}$ \\
Velocity of sound $c_s$ & $5.11 \times 10^5$ cm s$^{-1}$ \\
Electron/hole wave function size $d_{e,h}^{j}$  & $5$ nm \\
First QD bare energy $W_A$ & $1.34$ eV \\
Second QD bare energy $W_B$ & $1.38$ eV\\
Electron/hole cutoff frequency $\omega_{e,h,eh}$ & $1.05$ meV\\
Electron-phonon coupling strength in 3D $c_3$ & $0.46$ meV \\
Electron-phonon coupling strength in 2D $c_2$ & $8.7$ meV \\
Electron-phonon coupling strength in 1D $c_1$ & $103$ meV \\
Scaling energy $\omega_0$ & $1$ meV\\
Photon timescale $T^*$ & $1$ ns\\
\hline
\end{tabularx}\par
\bigskip
\end{minipage}

\section{Method} \label{sec:Method}

While a perturbative treatment of the radiation bath under the Born-Markov approximation is enough to accurately describe the effect of the radiation field, a range of theoretical methods have been developed to investigate the how phonon interactions affects the system dynamics. In the limit of weak coupling, one can do perturbative expansions of the QD-phonon coupling, resulting in master equation descriptions of both Markovian \cite{ramsay2010damping,ramsay2011effect,nazir2008photon,gauger2008robust} and non-Markovian \cite{machnikowski2004resonant,alicki2004optimal,mogilevtsev2008driving} nature, as well as correlation expansions \cite{krugel2005role,forstner2003phonon,krugel2006back}. The polaron transform \cite{mahan2000many} in conjunction with a perturbative expansion in the polaron-transformed basis can account for various non-perturbative effects not captured by the weak-coupling treatment \cite{wurger1998strong,wilson2002quantum,mccutcheon2010quantum}. Non-perturbative numerically exact techniques that rely on the calculation of the path integral have also been implemented \cite{vagov2007nonmonotonic, mccutcheon2011general}. 

Polaronic effects in the QD dynamics are smaller at lower temperatures ($T<30$~K) \cite{mccutcheon2010quantum} and weaker coupling, and, as we will show later in Figs.~\ref{fig:fig3} and \ref{fig:fig4}), the phonon-induced entanglement and correlations in photon emission are most pronounced at low temperatures. We are therefore able to treat both the radiation and phonon bath in a Born-Markov fashion, discussing future possible improvements on this approximation in our conclusion.
The general form of the resulting second order master equation for the reduced density matrix of a system $\rho$ is~\cite{timm08}:
\bea\label{A5}
\dot{\rho} (t) = - i [H_0, \rho(t)]  -  \int_0^{\infty} \di \tau\ \mathrm{tr_B}\left\{ \left[H_I,\left[ e^{i (H_0+ H_B) \tau } H_I e^{-i (H_0+ H_B) \tau } ,\rho(t) \otimes \rho_B\right]\right] \right\}.
 \eea
 with $H_0 (H_B)$ the system (bath) Hamiltonian and $H_I$ the system-bath interaction; $\rho_B$ is the time-independent bath density matrix.
 
 \begin{figure}[t!]
	\centering
	\includegraphics[]{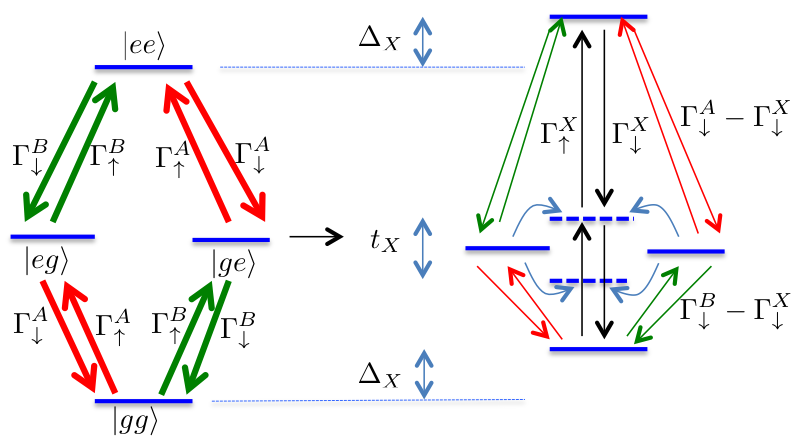}
	\caption{Schematic illustration of the effect of the common environment (ignoring single QD states renormalizations) corresponding to the master equation Eq. \ref{MasterEquation} in the case of $t_X,\Gamma_{\downarrow,\uparrow}^X>0$. In the left panel we present the schematic illustration of the dynamics when the phonons are modelled by two phonon baths interacting separately with the QDs while in the right panel we add the effects appearing due to the interaction with a common phonon bath. Red (green) arrows correspond to relaxation and pumping of the first (second) QD while black arrows correspond to relaxation and pumping involving the two QD entangled states $\ket{\psi^+}$, $\ket{\psi^-}$. The dashed states represent the entangled states $\ket{\psi^+}$ (upper state) and $\ket{\psi^-}$ (lower state) while the solid states represent the single QD states. Notice that there are three effects. Firstly, there is a shifting of the two QD states $\ket{gg}$ and $\ket{ee}$ by $\Delta_X$ which has no relevance when the steady state is reached. Secondly there is a splitting of the one excitation entangled states $\ket{\psi^+}$ and $\ket{\psi^-}$ by $t_X$, mediated by virtual phonons (the blue arrows show how these new system eigenstates form). Finally, collective phonon interactions result in pumping and relaxation processes involving the entangled states $\ket{\psi^+}$ and $\ket{\psi^-}$; these occur at rates $\Gamma_{\downarrow,\uparrow}^X$. Owing to this, the pumping and relaxation of the single QD states (represented by thick arrows on the left panel) decrease correspondingly by $\Gamma_{\downarrow,\uparrow}^X$ (represented by thinner arrows on the right panel).}
	\label{fig:figME}
\end{figure}

We can apply this general formulation to our specific case: It is straightforward to trace out the phonon and radiation reservoirs and obtain a master equation in Lindblad form (for details see \ref{app:AppendixA}). For the case of $3$D phonons we obtain: 
\bea\label{MasterEquation}
\dot{\rho}&=&-i \left[H'+H_X,\rho \right]+ \sum_j {\cal L}_j [\rho]+ {\cal L}_{X}[\rho],\\
H'&=& \frac{W_A'}{2} \sigma_z^A  + \frac{W_B'}{2} \sigma_z^B\nonumber ,\\
H_X&=&\frac{1}{2}t_X( \sigma_+^A \sigma_-^B + \sigma_-^A\sigma_+^B)+\frac{\Delta_X}{2}\sigma_z^A\sigma_z^B\nonumber, \\
{\cal L}_j [\rho]&=&(\Gamma_\downarrow^j-\abs{\Gamma_\downarrow^X}) {\cal D} (\sigma_-^j, \rho)+(\Gamma_\uparrow^j -\abs{\Gamma_\uparrow^X}){\cal D} (\sigma_+^j, \rho) \nonumber,\\
{\cal L}_X[\rho]&=&\left|\Gamma_\downarrow^X\right|{\cal D}(\sigma_-^A \pm \sigma_-^B, \rho)+\left|\Gamma_\uparrow^X\right|{\cal D}(\sigma_+^A \pm \sigma_+^B, \rho),\nonumber
\eea
where we define the dissipators ${\cal D} (x,\rho) \equiv  x \rho x^\dagger -1/2 \rho x^\dagger x -  1/2 x^\dagger x \rho$, and in ${\cal L}_X[\rho]$ the plus (minus) sign is chosen for positive (negative) rates. The various new parameters are defined as: 
\bea\label{rates}
W_j'&=&W_j+\Delta_j,\\
\Delta_j&=&\frac{\sin^2\theta_j}{2} {\cal P} \int_0^\infty \di \omega \frac{[2\bar{n}(\omega)+1]W_j}{\omega^2-W_j^2}\frac{J_p(\omega)}{2\pi} + \cos\theta_j{\cal P}\int_0^\infty \di \omega \frac{1}{\omega} \frac{J_p(\omega)+J_X(\omega) }{2\pi}\nonumber,\\
\Delta_X&=&\cos\theta_A\cos\theta_B{\cal P} \int_0^\infty \frac{1}{\omega} \frac{J_X(\omega)}{2\pi}\nonumber,\\
t_X&=&\frac{\sin\theta_A\sin\theta_B}{2}{\cal P}\int_0^\infty \left(\frac{1}{\omega^2-W_A^2}+\frac{1}{\omega^2-W_B^2}\right) \frac{\omega J_X(\omega)}{2\pi}\nonumber,\\
\Gamma_\downarrow^j &=&J_\gamma^j (\omega_l^j)\cos^4\frac{\theta_j}{2} +   \frac{\sin^2\theta_j}{4}J_p (W_j) [\bar{n}(W_j)+1]\nonumber,\\
\Gamma_\uparrow^j &=&\sin^4\frac{\theta_j}{2}+\frac{\sin^2\theta_j}{4}J_p (W_j) \bar{n}(W_j)\nonumber,\\
\Gamma_\phi^j&=&J_\gamma^j(\omega_l^j)\frac{\sin^2\theta_j}{4}\nonumber,\\
\Gamma_\downarrow^X&=& \frac{\sin\theta_A\sin\theta_B}{8} \left\{  J_X (W_A) [\bar{n}(W_A)+1] +J_X(W_B)[\bar{n}(W_B)+1]\right\}\nonumber,\\
\Gamma_\uparrow^X&=& \frac{\sin\theta_A\sin\theta_B }{8}  \left\{J_X (W_A) \bar{n}(W_A)+J_X (W_B) \bar{n}(W_B)\right\} \nonumber,
\eea
where, $\bar{n}(\omega)=1/(\exp[\omega/(k_B T)]-1)$ is the average phonon number at temperature $T$ and frequency $\omega$. In addition to the phonon and photon spectral density function we have defined the spectral density function of the common phonon field as: 
\bea
J_X(\omega) &\equiv&2\pi \sum_{\bm{k}} |t_{\bm{k}}|^2 e^{ i \bm{k}\cdot \bm{d}} \delta(\omega_{\bm{k}}-\omega) = F(\omega d/c_s)J_p(\omega),
\eea
where $c_s$ is the speed of sound in GaAs and $F(x)$ is a function that, in the case of 3D phonons, has the form $F(x)=\mathrm{sinc}(x)$.

\begin{figure}[t!]
    \centering
    \includegraphics[]{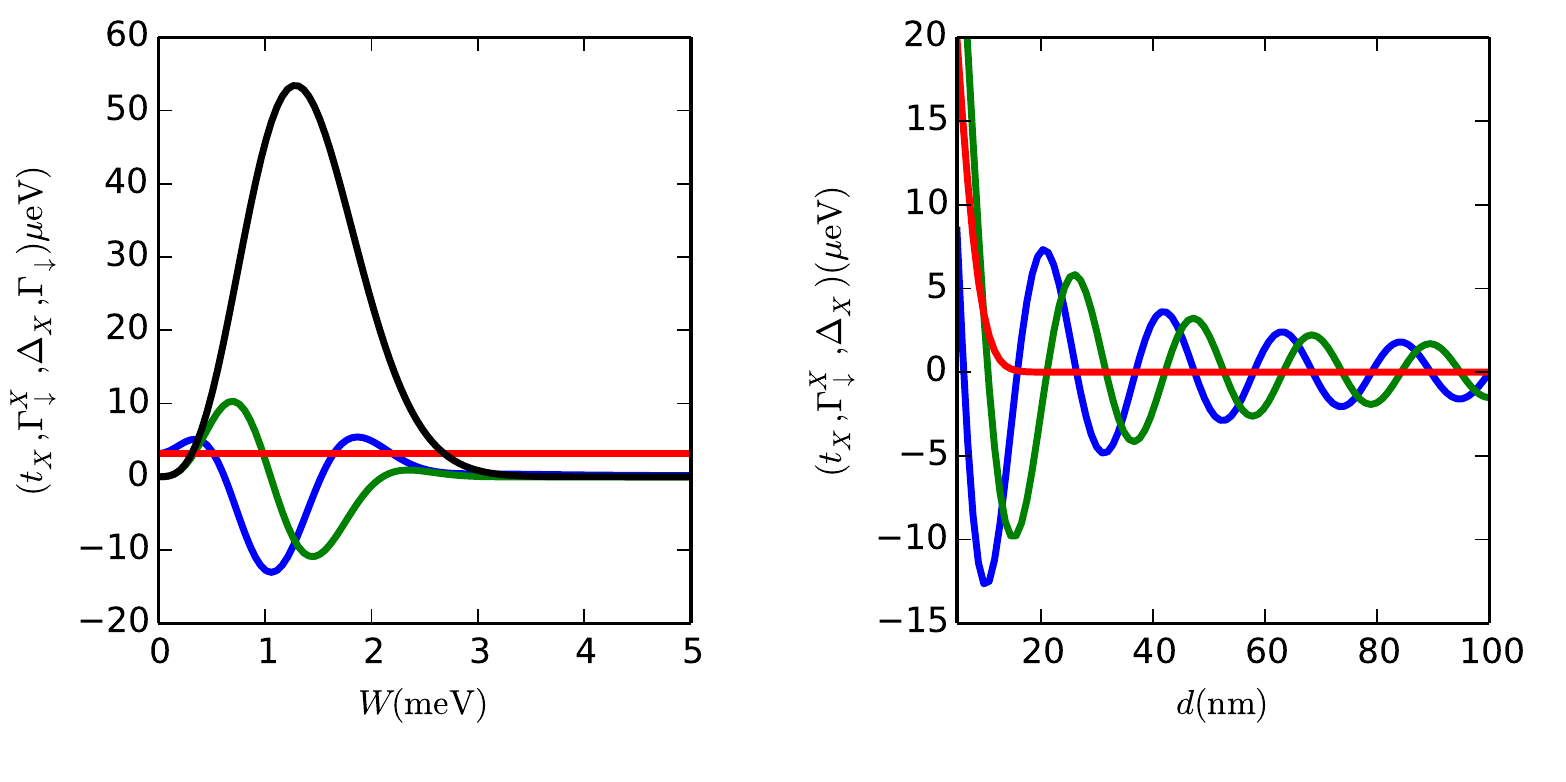}
    \caption{$t_X(W)$ (blue line), $\Gamma_\downarrow^X(W)$ (green line), $\Delta_X(W)$ (red line) and $\Gamma_\downarrow^{A,B}(W)$ (black) for varying $W=W_A=W_B$ (left panel) and varying distance between the dots $d$ (right panel). $W$ does not vary with $d$ so there is no black line on the right panel. Since these plots are meant only to give an idea of the strengths of the different couplings we evaluate $t_X$,$\Gamma_\downarrow^X$ and $\Gamma_\downarrow^{A,B}$  at $\theta_A=\theta_B=\pi/2$ and $\Delta_X$ at $\theta_A=\theta_B=0$. The values of the couplings are then calculated from Eq.~(\ref{rates}).}
    \label{fig:fig0}
\end{figure}

The master equation (\ref{MasterEquation}) is fully physical -- i.e. it is in Lindblad form with all rates positive -- as long as $\Gamma_{\downarrow,\uparrow}^X < \Gamma_{\downarrow}^j$. This is always the case since $F_D(x)\le1$. In the above equations we have isolated the effects that result solely due to the interaction of the QDs with a common phonon bath from the effects that would still appear even when the dots interact with separate phonon baths. We have denoted the corresponding Hamiltonian, dissipators and rates due to this common bath with a subscript $X$ since they only appear when there is a finite cross correlation between the local phonon environments of the two QDs. 

The effects that appear when the QDs interact with independent, separate phonon and photon baths are already well-known. The master equation for this case can be obtained from Eq.~\ref{MasterEquation} by setting $J_X(\omega)=0$, which corresponds physically to very distant or off resonant QDs. This leads first to a renormalization of the QD energies by $\Delta_j$, which has two contributions. One (whose size is proportional to $\sin^2\theta_j$) is most effective for resonantly driven dots and disappears for far-detuned driving lasers; the other (proportional to $\cos\theta_j$) is negligible in resonantly driven dots but is significant for far-detuned driving lasers. Independent interactions also lead to relaxation and pumping of each QD, at rates $\Gamma_\downarrow^j$ and $\Gamma_\uparrow^j$ respectively. This is caused by both phonons and photons, with the former being most effective when the dots are resonantly driven.
The photon contribution is more nuanced: for red-detuned driving lasers (i.e. $\theta=0$) the photons will induce relaxation, for blue-detuned driving lasers (i.e. $\theta=\pi$) they will induce pumping, while for resonantly driven QDs the photons will induce relaxation and pumping at equal rates. The photon bath will also induce pure dephasing of the QDs at rates $\Gamma_\phi^j$ that is most effective for resonantly driven dots, and which decays to zero for far-detuned driving lasers. 

When terms associated with the common phonon bath are introduced, this leads most straightforwardly to modified renormalization, relaxation and pumping rates for each dot individually. However, there are additional effects embedded in $H_X$ and ${\cal L}_X$ that lead to completely different physics. We present these additional effects schematically in Fig.~\ref{fig:figME} and discuss them in  detail below. 

The coherent term $H_X$, corresponding to processes mediated by virtual phonons, contains a renormalization and an elastic interaction between the two QDs. The renormalization part causes a shifting in energy of the eigenstates $\ket{gg}$ and $\ket{ee}$ by $\Delta_X$, while keeping the one excitation eigenstates $\ket{eg}$ and $\ket{ge}$ fixed. This does not affect the steady-state dynamics, so it is not important in our further calculations. By contrast, the induced elastic interaction between the two QDs, with strength $t_X$, changes the QD coherent dynamics and steady state properties. The one exciton states $\ket{eg}$ and $\ket{ge}$ become coupled elastically such that the single exciton eigenstates of the system are entangled states split by $t_X$. When the two excitons are resonant the new eigenstates are the usual symmetric and antisymmetric states: $\ket{\psi^+}=(\ket{eg}+\ket{ge})/\sqrt{2}$ and  $\ket{\psi^-}=(\ket{eg}-\ket{ge})/\sqrt{2}$. 


The incoherent term ${\cal L}_X$, corresponding to processes mediated by real phonons, leads to relaxation and pumping between the two QD states. When the qubits interact separFig.~ately with the phonon environment, the phonon field induces relaxation and pumping between the single QD basis states  $\ket{g}_j$ and $\ket{e}_j$ as we saw previously. Since these decoherence mechanisms affect single qubit states they will tend to destroy any coherence between the two QDs. However, the interaction with a common bath induces relaxation (accompanied by phonon emission) and pumping (accompanied by phonon absorption) between the two QD entangled states $\ket{\psi^+}$ and $\ket{\psi^{-}}$  and the two QD states $\ket{gg}$ and $\ket{ee}$ with rates $\Gamma_{\downarrow}^X$ (relaxation) and $\Gamma_{\uparrow}^X$ (pumping) -- see Fig.~\ref{fig:figME}. If the rate constants for these processes are favourable, then entangled steady states of the two QDs can result. 


The induced elastic coupling strength and the inter-QD pumping and relaxation rates are both proportional to $\sin\theta_A\sin\theta_B$ and are therefore most effective when the dots are resonantly driven.
To get an idea of the strength of these couplings we plot them in Fig.~\ref{fig:fig0}, for the case of 3D phonons. The couplings exhibit an oscillatory behaviour in both plots, due to the oscillatory function $F_{\mathds D}(x)$. The amplitude of the oscillations decreases with increasing distance between the dots due phonon dispersion.  

From the master equation we wish to obtain the equations of motion of the expectation values of system operators. 
We find that the resulting equations decouple into a set that includes all the population dynamics
(i.e. the density matrix elements $\rho_{ee-ee},\rho_{eg-eg},\rho_{ge-ge},\rho_{gg-gg}$ where $\rho_{x-y}\equiv \bra{x}\rho \ket{y}$), together with the one excitation coherences $\rho_{eg-ge}$ and $\rho_{ge-eg}$. All other coherences decouple and decay to zero in the long time limit, and so are not relevant for steady state calculations.
A convenient operator basis is therefore formed by the single qubit operators $\sigma_z^j$, the two qubit operator $\sigma_z^A \sigma_z^B$ and the two coherences, which in a rotating frame have the form $c (t)=\chi_r+i \chi_r$ where $\chi_r= \Re\{e^{i (W_B'-W_A') t } \bra{ge} \rho \ket{eg} \}$ and $\chi_i= \Im\{e^{i (W_B'-W_A') t }\bra{ge}\rho\ket{eg} \} $. 
In order to simplify the form of the resulting equations of motion we define 

\bea
\gamma_j=&\Gamma_\downarrow^j+\Gamma_\uparrow^j &\textrm{(single QD population difference relaxation rate),}\\
\gamma_d^j=&2 \Gamma_\phi^j + \gamma_j/2 &\textrm{(single QD dephasing rate),}\\
D_j=&(\Gamma_\uparrow^j - \Gamma_\downarrow^j)/\gamma_j &\textrm{(single QD relative steady state inversion),}\\
D=&D_A+D_B & \textrm{(total steady state inversion),}\\
\delta D=&D_A-D_B & \textrm{(inversion difference between QDs),}\\
\gamma=&\gamma_A+\gamma_B &\textrm{(total relaxation rate),}\\
\gamma_d=&\gamma_d^A+\gamma_d^B &\textrm{(total dephasing rate),}\\ 
\Gamma^X=&\Gamma_\downarrow^X+\Gamma_\uparrow^X& \textrm{(sum of cross relaxation rates),}\\ 
\delta \Gamma^X=&\Gamma_\downarrow^X-\Gamma_\uparrow^X & \textrm{(difference of cross relaxation rates).}
\eea
We may then write the following closed set of equations: 
\bea\label{EOM}
\dot{\bm{x}} &=&M \bm{x} +\bm{k}, \\
M &=& \left(\begin{matrix}

-\gamma_A	& 0 & 0 & -2\delta\Gamma^X&2 t_X \\

0&-\gamma_B  	& 0 &-2\delta\Gamma^X& -2 t_X \\

\gamma_B D_B & \gamma_A D_A & -\gamma & 4 \Gamma^X & 0\\

\delta\Gamma^X/4&\delta\Gamma^X/4 & \Gamma^X/2 &-\gamma_d &0\\

t_X/4&- t_X/4& 0&0&-\gamma_d\\
\end{matrix} \right),\\
\bm{x}^T &=&
\left(\begin{matrix}
 \langle \sigma_z^A \rangle &  \langle \sigma_z^B \rangle &   \langle \sigma_z^A \sigma_z^B \rangle & \chi_r &\chi_i  \end{matrix} \right),\\
 \bm{k}^T &=&
\left(\begin{matrix}
\gamma_A D_A &  \gamma_B D_B & 0   &  0&0 \end{matrix} \right).
\eea

\section{Steady-state}
\label{sec:Steady}
When a steady-state is reached, the above equations of motion can be solved exactly to obtain the steady state vector $\bm{x}_{ss}$. Although a full exact solution is straightforward, it is cumbersome and offers little insight. Therefore, in this section we will only consider the case when the phonon induced couplings $t_X$ and $\Gamma^X$ are small in comparison to the other decoherence rates in the problem (i.e. $t_X, \Gamma_X \ll \gamma_d$). In this pertubative limit we obtain solutions for the steady-state expectation values of the system operators $\bm{x}_{ss}$:
\bea \label{operators}
\langle \sigma_z^A \rangle_{ss} &=& D_A + \frac{2 t_X}{\gamma_A}\chi_i - \frac{2 \delta \Gamma^X}{\gamma_A} \chi_r\equiv D_A+\delta_A, \\
\langle \sigma_z^B \rangle_{ss} &=& D_B - \frac{2 t_X}{\gamma_B}\chi_i - \frac{2 \delta \Gamma^X}{\gamma_B} \chi_r\equiv D_B + \delta_B,\\
\langle \sigma_z^A\sigma_z^B \rangle_{ss} &=& \langle \sigma_z^A \rangle_{ss} \langle \sigma_z^B \rangle_{ss} -\delta_A\delta_B + \frac{2}{\gamma}\left( t_X \delta D \chi_i +\delta \Gamma^X D \chi_r+2\Gamma^X \chi_r\right), \\
\chi_r&=&\frac{\delta \Gamma^X}{4\gamma_d}\left(\langle \sigma_z^A\rangle_{ss}+\langle \sigma_z^B\rangle_{ss}\right)+\frac{\Gamma^X}{2\gamma_d}\langle\sigma_z^A\sigma_z^B \rangle_{ss}\approx\frac{\delta \Gamma^X}{4\gamma_d} D+\frac{\Gamma^X}{2\gamma_d}D_AD_B\label{cr}, \\
\chi_i&=&\frac{t_X}{4\gamma_d}\left(\langle \sigma_z^B \rangle_{ss} - \langle \sigma_z^A\rangle_{ss}\right)\approx-\frac{t_X}{4\gamma_d}\delta D\label{ci}.
\label{operatorslast}\eea
In fact, the expressions obtained for $\chi_{i,r}$ are correct up to second order in $(t_X,\Gamma^X)/\gamma_d$ while the expressions obtained for $\langle \sigma_z^j\rangle$ and $\langle \sigma_z^A\sigma_z^B\rangle_{ss}$ are correct to third order in $(t_X,\Gamma^X)/\gamma_d$ since there is no contribution to $\chi_{i,r}$ of second order in $(t_X,\Gamma^X)/\gamma_d$. 

The coherences $\chi_{i,r}$ depend strongly on the values of the $D_j$, the qubit inversion in the absence of any QD interaction. To obtain insight into the behaviour of these coherences we consider how the $D_j$ depend on the detuning angle $\theta_j$. At zero temperature we find: 
\bea
D_j(\theta_j)=\frac{\Gamma_\uparrow^j-\Gamma_\downarrow^j}{\Gamma_\uparrow^j+\Gamma_\downarrow^j}=\frac{-\cos\theta_j J_\gamma^j(\omega_l^j)-\frac{\sin^2\theta_j}{4}J_p(W_j)}{(1-\frac{\sin^2\theta_j}{2})J_\gamma^j(\omega_l^j) + \frac{\sin^2\theta_j}{4}J_p(W_j)}.
\eea
We can see from this that there are two potentially competing processes at work. First, the phonon coupling always tends to relax the QD to its ground state, regardless of $\theta$. However, the photon field will invert the QD when it is driven by a far blue-detuned laser (i.e. $D_j(\pi) =1$), while relaxing the QD to its ground state when it is driven by a far red-detuned laser (i.e. $D_j(0) =-1$). The strength of the radiation coupling then determines how blue-detuned the driving lasers must be to counteract the effect of the phonon field and invert the QD. We will need to use the appearance of such an inversion to explain some of our key results below.


Up to this point we have been completely general regarding the relationship between the angles $\theta_A$ and $\theta_B$ of the two QDs. However, in order to further investigate the strength of the coherences $\chi_{i,r}$ (and therefore the effects due to the induced interaction between the QDs) we need to be more specific about the relationship between the two driving lasers. Therefore, for the rest of the paper we will focus on two particular cases:
\begin{description}
\item[Case 1: Similarly detuned driving lasers.]
Firstly we consider the case when $ \theta_A=\theta_B$ and the two QDs exhibit the same behaviour. In this case $D_A=D_B$ and therefore $\chi_i=0$. This regime is very useful, since then all the terms involving the phonon induced coherent coupling $t_X$ are also zero in Eqs.~\ref{operators} to \ref{operatorslast} and so we can isolate the incoherent QD coupling mediated by real phonons. We should also note that, at zero temperature, $\chi_r=0$ for far red-detuned driving lasers (i.e. $\chi_r(\theta= 0)=0$) and therefore we should use blue-detuned driving lasers to ensure a contribution of $\chi_r$.
\begin{figure}[t!]
	\centering
	\includegraphics[]{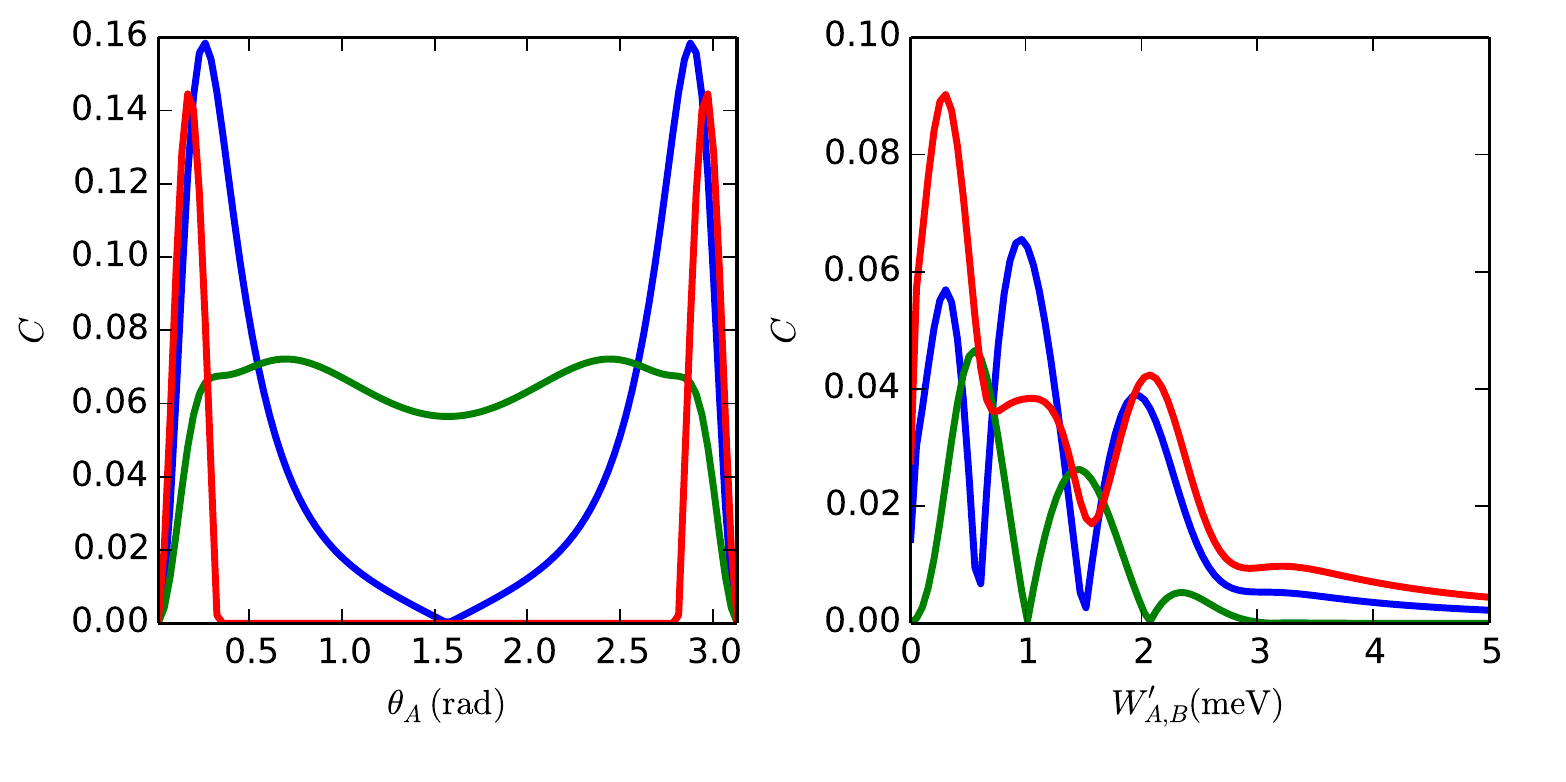}
	\caption{Steady-state concurrence $C$ (red line), $\chi_i$ (blue line) and $\chi_r$ (green line)  for varying driving laser angle $\theta_A$ for a fixed QD frequency $W_A'=W_B'=0.3$~meV (left panel) and for varying QD frequency with a fixed first driving laser angle $\theta=\pi-0.1$ (right panel).  The plots are obtained from the full numerical solution of the master equation Eq.~\ref{GeneralMaster} but agree well with the analytical formulas 
Eqs. \ref{cr}, \ref{ci} and \ref{concurrence}.}
	\label{fig:fig1}
\end{figure}
\item[Case 2: Oppositely detuned driving lasers.]
Secondly we consider the case $\theta_A = \pi- \theta_B$ which will allow us to investigate the coherent QD coupling mediated by virtual phonons. In this case, for far off resonant driving fields one QD will be in the excited state and the other in the ground state such that $D_A=-D_B$, resulting in a finite $\chi_i$ and $\chi_r$. 

\end{description}

Since the coupling to the same phonon field results non-zero steady-state coherence between the two qubit states $\ket{eg}$ and $\ket{ge}$ we expect that the two QDs will actually be entangled through their interaction with the phonon bath. For a bipartite system the most common measure of entanglement between the two subsystems is the concurrence, defined as $C(\rho) = \max\left\{0,\lambda_1-\lambda_2-\lambda_3-\lambda_4\right\}$ where $\lambda_i$ are the square roots of the eigenvalues of the matrix $\rho \tilde{\rho}$  listed in decreasing order where $\tilde{\rho } = (\sigma_y \otimes \sigma_y) \rho^* \left( \sigma_y \otimes \sigma_y\right)$ \cite{wootters1998entanglement}. In our case the concurrence has the following expresion: 
\bea\label{concurrence}
C&=&2\cdot \max\left\{0,\min \left[\abs{\rho_{eg-ge}}, \sqrt{\rho_{eg-eg} \rho_{ge-ge}}  \right] -  \sqrt{\rho_{ee-ee}\rho_{gg-gg}} \right\}\label{concurrence1},\\
C&=&\max\left\{0,\min \left[2 \sqrt{\chi_i^2+\chi_r^2},\frac{1}{2} \sqrt{(1-\langle \sigma_z^A\sigma_z^B \rangle_{ss})^2-(\langle \sigma_z^A \rangle_{ss}-\langle\sigma_z^B\rangle_{ss})^2}  \right] \right.\nonumber\\
& &\left. \quad \quad \quad- \frac{1}{2} \sqrt{(1+\langle \sigma_z^A\sigma_z^B \rangle_{ss})^2-(\langle \sigma_z^A \rangle_{ss}+\langle\sigma_z^B\rangle_{ss})^2} \right\}\nonumber.
\eea
The case of oppositely detuned driving lasers is a lot more efficient in entangling the QDs since it includes contributions from both $\chi_r$ and $\chi_i$. We plot concurrence for this case in Fig.~\ref{fig:fig1}, as a function of the driving angle $\theta$ for a fixed Rabi frequency $W=0.3$~meV (left panel) and as a function of the Rabi frequency strength $W$ for a fixed angle $\theta=0.1$~rad (right panel), both parameter regimes easily accessible experimentally. We can see that the entanglement is large far from resonance because that is where the coherences $\chi_i,\chi_r$ are largest. We also notice that both the virtual phonon coupling and the real phonon coupling give rise to entanglement. 

\section{Photon statistics} \label{sec:Photon}
In obtaining the master equation for the reduced density matrix we lost track of the emitted field by tracing out the photon bath operators $a_{\bm{q}}$. However, correlation measurements of the emitted photon fields can yield valuable information about the QD properties and the phonon bath \cite{brown1956correlation,gerardot2005photon}. In this section we show how these correlation measurements can be used to to find signatures of a common phonon bath and to measure the strength of the induced couplings $t_X$ and $\Gamma^X$. 

An experimentally accessible quantity which is very sensitive to QD coupling mechanisms is the (normalized) intensity cross correlation function. For two photon modes $a_j$ and $a_k$ the intensity cross correlation function is defined as:
\bea
g_{jk}^{(2)} (\tau)=\frac{ \langle a_j^\dagger (0) a_k^\dagger(\tau) a_k(\tau) a_j(0) \rangle }{ \langle a_j^\dagger (0) a_j \rangle \langle a_k^\dagger(0) a_k(0)  \rangle},
\eea
$g_{jk}^{(2)}(\tau)$ is proportional to the probability of detecting a photon in mode $k$ at time $t=\tau$ given that a photon in mode $j$ was detected at time $t=0$.  If the two photons are completely uncorrelated the intensity correlation function is $1$; any deviation from $1$ is a signature of correlated photons, and so also a signature of correlations in the QD states which led to the photon emission. 


In order to measure $g_{jk}^{(2)}$ experimentally one can use a Hanbury Brown-Twiss interferometer. In this set-up the two photon streams emitted by the QDs  are detected separately using a beam splitter and corresponding frequency filters. The photon detectors ${\cal A}$ and ${\cal B}$ used to detect the photon streams are connected to a timer which is activated when a photon is detected by detector ${\cal A}$ and is stopped when another photon is detected by detector ${\cal B}$ (negative time correlations $g_{jk}^{(2)}(-\tau)$ can be obtained by delaying the second photon stream by $\tau$) . The delay time between the two detection events is recorded and a delay-histogram can be obtained from which $g_{jk}^{(2)}(\tau)$ can be obtained. Since the timing resolution of the detectors is not perfect, the actual measured function will be:
\bea
\tilde{g}_{jk}^{(2)}(\tau)=\int_{-\infty}^\infty \di t \frac{1}{\sigma \sqrt{2 \pi}} e^{-\frac{1}{2}\left( \frac{t-\tau}{\sigma}  \right)^2} g_{jk}^{(2)}(t),
\eea
where $\sigma$ captures the timing jitter of the photon detectors. Here we will take $\sigma=150$~ps, which is achievable with state-of-the-art photon detectors \cite{marsili2013detecting,eisaman2011invited}.

\begin{figure}[t!]
	\centering
	\includegraphics[]{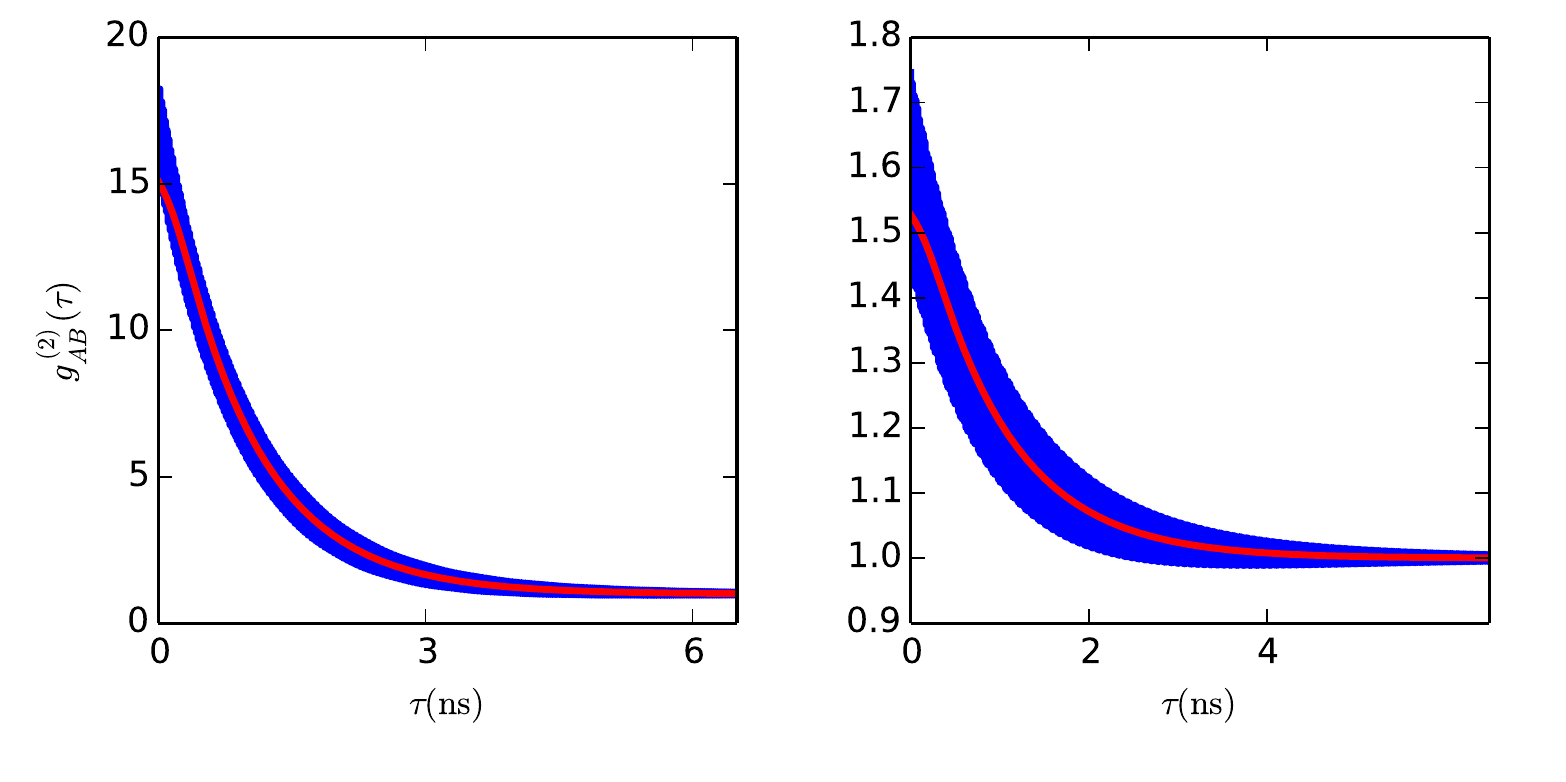}
	\caption{$g_{AB}^{(2)}(\tau)$ (blue line) and $\tilde{g}_{AB}^{(2)}(\tau)$(red line) for the case of oppositely detuned driving lasers (left panel) and for the case of similarly detuned lasers (right panel) . All parameters are from Table \ref{tab:parameters} while $W_A'=W_B'=0.2$ and $\theta_A=\pi-0.1$. The plots are obtained by numerically solving the master equation \ref{GeneralMaster}.}
	\label{fig:fig2}
\end{figure}

In order to relate the correlations in the emitted photon fields to the QD operators and thus to the phonon induced couplings we use the input-output formalism~\cite{walls08}. We start from the initial Hamiltonian Eq.~\ref{StartingHamiltonian}. According to the input-output formalism, the input field driving the QDs and the output field emitted by the QDs are related through the relation $a_{out} (t) = a_{in} (t) + \sqrt{\Gamma_A} c_A(t ) + \sqrt{\Gamma_B} c_B(t )  $, where $\Gamma_j = J_\gamma (\omega_l^j) [1 + \bar{n}_\gamma (\omega_l^j,T)] $ and for simplicity's sake we make the reasonable approximation $J_\gamma(\omega_l^j \pm W_j)=J_\gamma(\omega_l^j)$. $a_{in}$ is given by the classical driving field amplitude corresponding to the two lasers incident on the QDs: in a typical experimental set-up this contribution is eliminated. In terms of the slowly rotating operators $\tilde{c}_j$ the output field is then $a_{out}(t) = \sqrt{\Gamma_A} \exp( i \omega_l^A t) \tilde{c}_A + \sqrt{\Gamma_B} \exp (i \omega_l^B t) \tilde{c}_B$. Thus we can see that the output field is composed of two separate photon fields, one emitted by the first QD with frequencies centered around $\omega_l^A$ and one emitted by the second QD with frequencies centered around $\omega_l^B$. Therefore, so long as the dot emission spectra are well resolved, a simple grating can be used to separate the output field into $a_A (t) = \sqrt{\Gamma_A} c_A$ and $a_B (t) = \sqrt{\Gamma_B} c_B$. These two fields are then incident on detectors ${\cal A}$ and ${\cal B}$ respectively. Therefore, according to input-output formalism: 
\bea
g_{AB}^{(2)} (\tau)= \frac{ \left\langle c_A^\dagger (0) c_B^\dagger(\tau) c_B(\tau) c_A(0) \right\rangle }{\left\langle c_A^\dagger(0) c_A(0) \right\rangle \left\langle c_B^\dagger(0) c_B(0)\right\rangle }.
\eea
In order to evaluate this function we need to solve the time dynamics from the equations of motion, Eq.~\ref{EOM} and then use the Quantum Regression Theorem~\cite{breuer02} to relate the correlation functions to the QD operators. The time dynamics solution requires obtaining the eigenvalues of the matrix $M$ and they can only be obtained numerically. Therefore, we are able to obtain $g_{AB}^{(2)}(\tau) $ only numerically. However, $g_{AB}^{(2)}(0)$ can be evaluated analytically since it only requires the evaluation of the two QD operator $c_A^\dagger c_A c_B^\dagger c_B$ in the steady state: 
\bea
\langle c_A^\dagger c_Ac_B^\dagger c_B\rangle_{ss}&=&\bra{XX} \rho_{ss} \ket{XX}=\frac{\sin(\theta_A)\sin(\theta_B)}{2} \chi_r\nonumber\\
 & &+\frac{1}{4}\left[1+\cos(\theta_A) \langle \sigma_z^A \rangle_{ss}+\cos(\theta_B)\langle \sigma_z^B\rangle_{ss}+\cos(\theta_A)\cos(\theta_B)\langle  \sigma_z^A\sigma_z^B\rangle_{ss} \right],\\
 \langle c_A^\dagger c_A  \rangle_{ss}&=&\bra{XX}\rho_{ss}\ket{XX}+\bra{X0}\rho_{ss}\ket{X0}=\frac{1}{2}\left[1+\cos(\theta_A) \langle\sigma_z^A \rangle_{ss} \right].
 \eea
 
\begin{figure}[t!]
	\centering
	\includegraphics[]{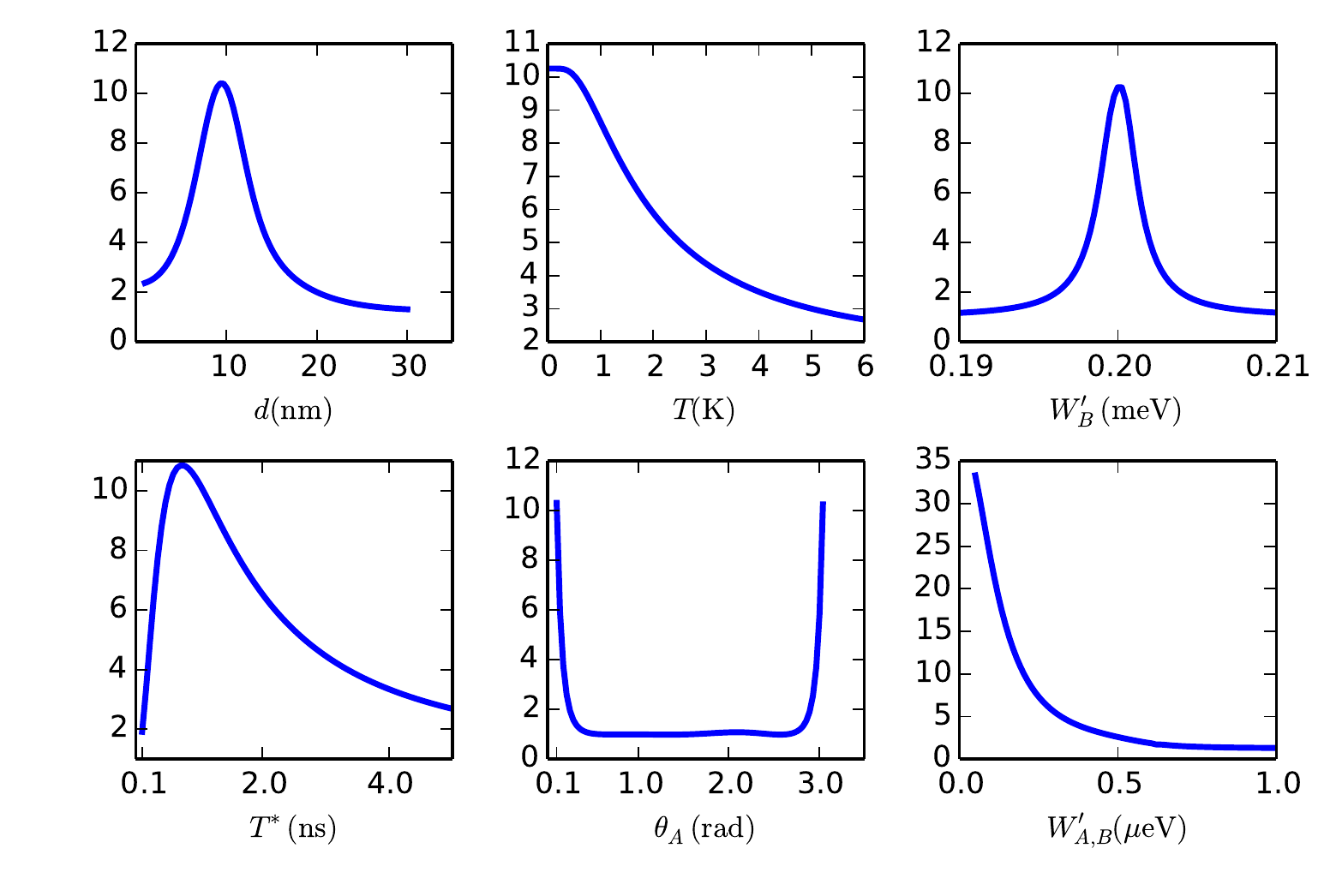}
	\caption{The common environment signature $\tilde{g}_{AB}^{(2)}(0)$ across different parameter regimes for the case of oppositely detuned lasers. We look at the dependence of the signature on distance $d$ between the dots (top left), temperature $T$ (top middle), second QD renormalized frequency $W_B'$ (top right), radiation field coupling strength $T^*$ (bottom left), first QD driving laser angle $\theta_A$ (bottom middle) and QD renormalized frequency $W_A'=W_B'$ (bottom right). Except for the parameters varied, in each panel the rest of the parameters are from Table \ref{tab:parameters} while $W_A'=W_B'=0.2$ and $\theta_A=\pi-\theta_B=\pi-0.1$. The plots are obtained by numerically solving the master equation \ref{GeneralMaster} (after including the radiation-field coupling effect as in Eq.~\ref{MasterEquation}).}
	\label{fig:fig3}
\end{figure}

From the above relations we obtain $g_{AB}^{(2)}(0)$: 
\bea\label{g20exact}
g_{AB}^{(2)}(0)&=&1+\frac{\cos(\theta_A)\cos(\theta_B) \left[\langle \sigma_z^A \sigma_z^B\rangle_{ss} - \langle\sigma_z^A \rangle_{ss} \langle\sigma_z^B \rangle_{ss}\right] +2\sin(\theta_A)\sin(\theta_B) \chi_r }{(1+\cos(\theta_A)\langle \sigma_z^A\rangle_{ss})(1+\cos(\theta_B)\langle \sigma_z^B \rangle)},\\
g_{AB}^{(2)}(0)&\approx& 1 + 2\frac{ \frac{\cos(\theta_A)\cos(\theta_B)}{\gamma}\left[  t_X \delta D \chi_i +(2\Gamma^X+\delta \Gamma^X D)\chi_r\right]+\sin(\theta_A)\sin(\theta_B)\chi_r}{(1+\cos(\theta_A)(D_A+\delta_A))(1+\cos(\theta_B)(D_B+\delta_B))} .
\eea
We can see that the incoherent coupling mediated by real phonons has both a first order and a second order (in the small parameters $t_X/\gamma,\Gamma_X^{\uparrow,\downarrow}/\gamma$) contribution to $g_{AB}^{(2)}(0)$ while the coherent coupling mediated by real phonons has only a second order contribution to $g_{AB}^{(2)}(0)$. However, the first order contribution  is most effective at resonance while the second order contribution is most effective far from resonance. Since the denominator of Eq.~\ref{g20exact} decreases at a faster rate than the coupling strengths $t_X,\Gamma^X$ we expect that we will see the largest effects far from resonance, even though these will be due to the second order process. 
\begin{figure}[h!]
	\centering
	\includegraphics[]{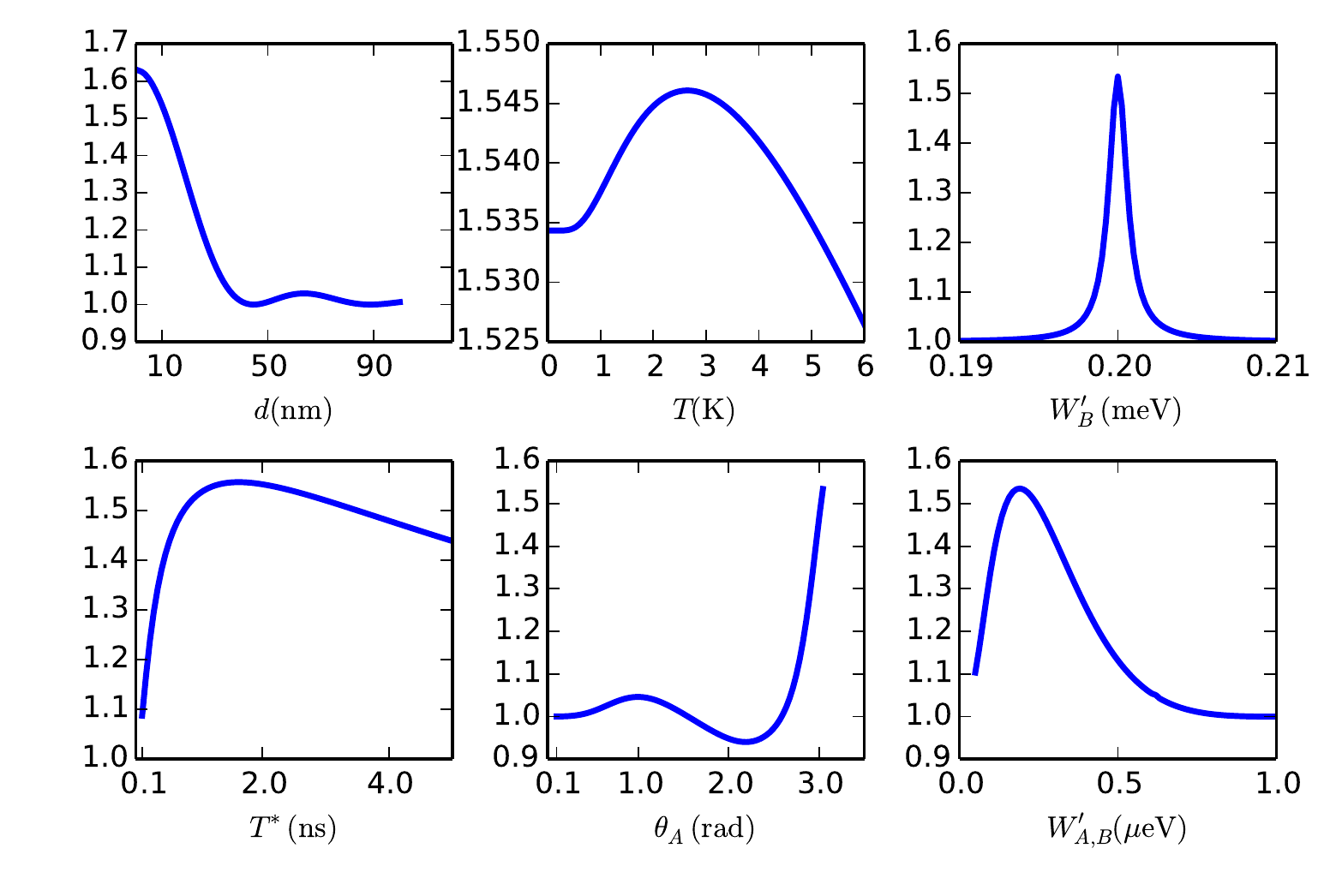}
	\caption{The common environment signature $\tilde{g}_{AB}^{(2)}(0)$ across different parameter regimes for the case of similarly detuned lasers. We look at the dependence of the signature on distance between the dots $d$ (top left), temperature $T$ (top middle), second QD renormalized frequency $W_B'$ (top right), radiation field coupling strength $T^*$ (bottom left), first QD driving laser angle $\theta_A$ (bottom middle) and QD renormalized frequency $W_A'=W_B'$ (bottom right).Except for the parameters varied, in each panel the rest of the parameters are from Table \ref{tab:parameters} while $W_A'=W_B'=0.2$ and $\theta_A=\theta_B=\pi-0.1$. The plots are obtained by numerically solving the master equation \ref{GeneralMaster} (after including the radiation-field coupling effect as in Eq.~\ref{MasterEquation}).}
	\label{fig:fig4}
\end{figure}
It is therefore possible to probe and characterise the common phonon environment by measuring $g^{(2)}(0)$. In the case of similarly detuned lasers, when there is no virtual phonon contribution, the real phonon coupling strength can be obtained directly from $g_{AB}^{(2)}(0)$. Then, by moving to the case of oppositely detuned lasers and inserting the known value of real phonon coupling, the virtual phonon coupling strength can also be found. 

In order to establish whether this simple signature can be observed using current technology, we take into account the finite smearing of the measurement due to timing jitter. We then only have access to the smeared function $\tilde{g}_{AB}^{(2)}(\tau)$, which can only be found numerically. $\tilde{g}_{AB}^{(2)}(\tau)$ is very sensitive to the driving lasers' mixing angles $\theta_{A,B}$ and renormalized QD frequency $W_{A,B}'$, so first, in Fig.~\ref{fig:fig2}, we take values for these parameters where the effect of the common phonon bath is most clearly seen. 
We show both an example of similarly detuned (right panel) and oppositely detuned lasers (left panel). We drive the first QD with a blue detuned laser ($\theta_A=\pi-0.1$) and use Rabi frequencies $W_A'=W_B'=0.2$meV and look at both $g_{AB}^{(2)}(\tau)$ (blue line) and $\tilde{g}_{AB}^{(2)}(\tau)$ (red line). The common phonon environment has a large impact on $\tilde{g}_{AB}^{(2)}(\tau)$, easily detectable experimentally. Indeed, in the case of oppositely detuned lasers the signature of the common phonon bath is extremely large. It would certainly be possible to use such data to fit for $t_X$ and $\Gamma^X$. 

We next show how these signatures change as several parameters are varied for the case of similarly (oppositely) detuned lasers in Fig.~\ref{fig:fig3} (Fig.~\ref{fig:fig4}). There are several differences between these two figures. For example, in the case of oppositely detuned lasers the largest signature is when the dot distance is around $10$nm (top left panel), a distance easily accessible experimentally. In contrast, for the case of similarly detuned lasers the signature is largest at zero dot separation, which is obviously unfeasible, but the signature is still large at $10$nm (top left panel). As expected, in the case of oppositely detuned lasers, we can see that the largest signature is at zero temperature (top middle panel) because $t_X$ is independent of temperature while the other decoherence rates grow with increasing temperature. However, because $\Gamma^X$ grows with increasing temperature, in the case of similarly detuned lasers there is an ideal temperature when the signature is largest  (top middle panel). We also see that, in the case of oppositely detuned lasers the largest signature is obtained when the QD frequencies  $W_{A,B}'$ are smallest while in the case of similarly detuned lasers there a finite $W_{A,B}'$ is optimal (bottom right panel).

There are also similarities between the signatures in Fig.~\ref{fig:fig3} and \ref{fig:fig4}: Both are largest when the two QDs are resonant with each other $W_A'=W_B'$ (top right panel) and the largest signature is for off resonant driving lasers (bottom middle panel). We also see that there are ideal radiation field coupling strengths (bottom left panel) at which the signatures are largest.

\section{Discussion and conclusion}\label{sec:Conclusion}

Using a weak coupling master equation approach, we have shown that two nominally un-coupled driven QDs are nevertheless effectively coupled through the interaction with a common environment. Using a second order Born-Markov master equation we isolated the effects that are solely due to the interaction of the QDs with the same phonon field, and which cannot be explained by considering the QDs coupled to separate phonon fields. We showed that the interaction with the phonon bath results in two types of induced interaction between the two QDs: an elastic interaction mediated by virtual phonons and an inelastic interaction mediated by real phonons. Both types of  interaction entangle the two QDs, and we have shown that photon statistics measurements can be used to obtain a signature of this common environment.


We have based our study on a weak coupling master equation. Our neglect of non-Markovian effects here will cause most deviation when the correlation time of the bath is of order the system dynamics timescale - i.e. when the QD frequencies are of the same order as the phonon bath cutoff frequencies $\omega_{e,h,eh}\approx 1$ meV. Therefore, our weak coupling predictions are quite reliable for the regime in which $W_{A,B}$ is smaller than this 1~meV scale. In future, we would like to extend this work beyond this regime by either introducing unitary transformations such as the polaron transformation~\cite{mccutcheon2010quantum}, or by using an exact approach such as QUAPI~\cite{MakriMakarov1995_1,MakriMakarov1995_2, mccutcheon2010quantum}.

\section{Acknowledgements}\label{sec:acknowledgements}

We would like to thank Jonathan Keeling, Paul Eastham, Helen Cammack, Ahsan Nazir and Jacob Iles-Smith for useful discussions. BWL thanks the Royal Society for a University Research Fellowship, OC would also like to thank the Royal Society for support and the School of Physics and Astronomy in St Andrews for kind hospitality.

\appendix
\section{Master equation for two systems interacting with the same reservoir bath}\label{app:AppendixA}
In this appendix we obtain a second order Born-Markov master equation for the general case of two systems coupled to the same reservoir. We start from the most general Hamiltonian: 
\bea
H&=&H_0+H_B+H_I,\\
H_B&=&\omega_{\bm{k}} c_{\bm{k}}^\dagger c_{\bm{k}},\\
H_I&=&\sum_{\bm{k}} (c_{\bm{k}} +c_{\bm{k}}^\dagger)( t_{\bm{k}} S_A  + g_{\bm{k}} S_B +h.c.),
\eea
where $H_0$ is the Hamiltonian of the two systems which can also contain interactions with other reservoirs and coupling between the two systems, $H_B$ is the Hamiltonian of the reservoir, and $H_I$ is the interaction Hamiltonian between the reservoir and the two systems. $S_A$ and $S_B$ can be any system operators describing the coupling to the reservoir. 

\subsection{Simple system-reservoir coupling operator}

Before deriving a general result, we first consider a simple case when $S_A=a+a^\dagger$ and $S_B=b+b^\dagger$ where $a=\ket{\psi_i^A}\langle\psi_j^A|$ and $b=\ket{\psi_k^B}\bra{\psi_l^B}$, and  where $\ket{\psi_{i/j}^A}$ and $\ket{\psi_{k/l}^B}$ are eigenstates of the system Hamiltonian $H_0$. In this case we have the relation $\exp(i H_0 t) a \exp(-iH_0 t)=\exp(-i \omega_A t) a$ and 
$\exp(i H_0 t) b \exp(-iH_0 t)=\exp(-i \omega_B t) b$, with $\omega_A$ and $\omega_B$ the relevant differences in eigenstate energies for systems $A$ and $B$.

In the Schr\"odinger picture the Born-Markov second-order master equation takes the form~\cite{timm08}:
\bea\label{rrr}
\dot{\rho} (t) = - i [H_0, \rho(t)]  -  \int_0^{\infty} \di \tau\ \mathrm{tr_B}\left\{ \left[H_I,\left[ e^{i (H_0+ H_B) \tau } H_I e^{-i (H_0+ H_B) \tau } ,\rho(t) \otimes \rho_B\right]\right] \right\}.
\eea

Assuming a bath in thermal equilibrium we trace out the bath operators to obtain the second order master equation:
\bea
\dot{\rho}&=&-i[ H_0+\Delta_1^A [a^\dagger,a]-\Delta_2^A \{a^\dagger,a\}+\Delta_1^B [b^\dagger,b]-\Delta_2^B \{b^\dagger,b\},\rho]\\
& & +\Gamma_\downarrow^A {\cal D} (a, \rho)+\Gamma_\uparrow^A {\cal D} (a^\dagger, \rho) + \Gamma_\downarrow^B {\cal D} (b, \rho)+\Gamma_\uparrow^B {\cal D} (b^\dagger, \rho)\nonumber\\
& &+{\cal D}(\Gamma_{\downarrow X}^{A}a+\Gamma_{\uparrow X}^{A}a^\dagger,b+b^\dagger, \rho)+{\cal D}(\Gamma_{\downarrow X}^B b+\Gamma_{\uparrow X}^B b^\dagger,a+a^\dagger, \rho),
\eea
where we have defined the dissipators ${\cal D}$ as:
\bea
{\cal D} (x,y, \rho)& \equiv &\frac{1}{2}\left[ x \rho y^\dagger +y \rho x^\dagger -y^\dagger x\rho-  \rho x^\dagger y  \right],\\
{\cal D} (x, \rho)& \equiv &{\cal D}(x,x, \rho),
\eea
and where: 
\bea
\Delta_1^j &=&{\cal P}\int_0^{\infty}\di \omega \ \frac{[2 \bar{n} (\omega_j) +1]\omega_j}{(\omega^2 -\omega_j^2)}\frac{J_j(\omega)}{2\pi},\\
\Delta_2^j &=&{\cal P}\int_0^{\infty}\di \omega \ \frac{\omega}{(\omega^2 -\omega_j^2)}\frac{J_j(\omega)}{2\pi},\\
\Gamma_\downarrow^j& =& J_j(\omega_j) [\bar{n}(\omega_j)+1] ,\\
\Gamma_\uparrow^j& =&J_j(\omega_j) \bar{n}(\omega_j),\\
\Gamma_{\downarrow X}^{j}& =&J_{X}^j(\omega_j) [\bar{n}(\omega_j)+1]+ i \cdot 2 {\cal P}\int_0^{\infty}\di \omega \left\{\frac{\bar{n}(\omega)+1}{\omega-\omega_j}\frac{J_X^j(\omega)}{2\pi}-\frac{\bar{n}(\omega)}{\omega+\omega_j}\frac{[J_X^j(\omega)]^*}{2\pi}\right\},\\
\Gamma_{\uparrow X}^{j}& =&[J_{X}^j(\omega_j)]^* \bar{n}(\omega_j)- i \cdot 2 {\cal P}\int_0^{\infty}\di \omega \left\{\frac{\bar{n}(\omega)}{\omega-\omega_j}\frac{[J_X^j(\omega)]^*}{2\pi}-\frac{\bar{n}(\omega)+1}{\omega+\omega_j}\frac{J_X^j(\omega)}{2\pi}\right\},
\eea
where, as in the main text,  $j\in\{A,B\}$, $\bar{n}(\omega)$ is the average phonon number of frequency $\omega$ at temperature $T$ and we have defined the following spectral density functions:
\bea
J_A(\omega) &=&2\pi \sum_{\bm{k}} \abs{t_{\bm{k}}}^2 \delta(\omega_{\bm{k}}-\omega),\\
J_B(\omega) &=&2\pi\sum_{\bm{k}} \abs{g_{\bm{k}}}^2 \delta(\omega_{\bm{k}}-\omega),\\
J_X^A(\omega) &=&2\pi \sum_{\bm{k}} t_{\bm{k}}^* g_{\bm{k}} \delta(\omega_{\bm{k}}-\omega),\\
J_X^B(\omega)&=&2\pi \sum_{\bm{k}} t_{\bm{k}} g_{\bm{k}}^* \delta(\omega_{\bm{k}}-\omega)=[J_X^A(\omega)]^*.
\eea
\subsection{Most general system-reservoir coupling}

In the previous section we have obtained the master equation in the case when the system reservoir coupling operator has a simple form $S_A=a + a^\dagger$ and $S_B=b+b^\dagger$ where $a$ is a transition between two eigenstates of the first QD and $b$ represents a transition between two eigenstates of the second QD. However, in general, the system reservoir coupling operator $S_j$ can be any operator acting on QD $j$. Since any operator on QD $j$ can be written as a sum of transitions between two eigenstates for that QD, then we can easily generalize the calculations in the previous question to obtain the master equation for a more general system reservoir coupling operator. 
 
The most general system-reservoir coupling operator can be written as $S_j=\sum_{q} \sigma_{qj} $ where $\sigma_{qj}=|\psi_{n}^j\rangle \langle \psi_m^j|$ where $|\psi_{n,m}\rangle$ are eigenstates of the $j$th system Hamiltonian; $q$ represents any pair of indices $\{n, m\}$. Therefore $\exp(i H_0 t) \sigma_{qj} \exp(-iH_0 t)=\exp(-i \omega_{qj} t) a$, and here $\omega_q$ is the relevant difference in eigenstate energies for the particular pair of indices $\{n, m\}$. Therefore, the most general Hamiltonian for the system-reservoir interaction can be written as: 
\bea
H&=&H_0+H_B+H_I,\\
H_B&=&\omega_{\bm{k}} c_{\bm{k}}^\dagger c_{\bm{k}},\\
H_I&=&\sum_{\bm{k}} \sum_{qj}  (c_{\bm{k}} +c_{\bm{k}}^\dagger) [t_{\bm{k}}^j \sigma_{qj} +h.c.].
\eea
After tracing out the reservoir in Born-Markov fashion we obtain the second order master equation: 
\bea
\dot{\rho}&=&-i\left[ H_0+\sum_{qj}\Delta_1^{qj} [\sigma_{qj}^\dagger,\sigma_{qj}]-\Delta_2^{qj} \{\sigma_{qj}^\dagger,\sigma_{qj}\},\rho\right]\\
& &+ \sum_{q,j} \left\{ \Gamma_\downarrow^{qj} {\cal D} (\sigma_{qj}, \rho)+\Gamma_\uparrow^{qj} {\cal D} (\sigma_{qj}^\dagger, \rho)\right\}+ \sum_{q,p\neq q,j}{\cal D}\left(\Gamma_{\downarrow x}^{qj} \sigma_{qj} +\Gamma_{\uparrow x}^{qj}\sigma_{qj}^\dagger  ,\sigma_{pj}+\sigma_{pj}^\dagger, \rho\right) \nonumber\\
& &+ \sum_{q,p}\left\{{\cal D}\left(\Gamma_{\downarrow X}^{qA}\sigma_{qA} + \Gamma_{\uparrow X}^{qA} \sigma_{qA}^\dagger,\sigma_{pB}+\sigma_{pB}^\dagger, \rho\right)+{\cal D}\left(\Gamma_{\downarrow X}^{pB}\sigma_{pB} + \Gamma_{\uparrow X}^{pB} \sigma_{pB}^\dagger,\sigma_{qA}+\sigma_{qA}^\dagger, \rho\right) \right\}\nonumber.
\eea
where 
\bea
\Delta_1^{qj} &=&{\cal P}\int_0^{\infty}\di \omega \ \frac{[2 \bar{n} (\omega_{qj}) +1]\omega_{qj}}{(\omega^2 -\omega_{qj}^2)}\frac{J_j(\omega)}{2\pi},\\
\Delta_2^{{qj}} &=&{\cal P}\int_0^{\infty}\di \omega \ \frac{\omega}{(\omega^2 -\omega_{qj}^2)}\frac{J_j(\omega)}{2\pi},\\
\Gamma_\downarrow^{qj}& =& J_j(\omega_{qj}) [\bar{n}(\omega_{qj})+1] ,\\
\Gamma_\uparrow^{qj}& =&J_{j}(\omega_{qj}) \bar{n}(\omega_{qj}),\\
\Gamma_{\downarrow x}^{{qj}}& =&J^j(\omega_{qj}) [\bar{n}(\omega_{qj})+1]+ i \cdot 2 {\cal P}\int_0^{\infty}\di \omega \left\{\frac{\bar{n}(\omega)+1}{\omega-\omega_{qj}}\frac{J^{j}(\omega)}{2\pi}-\frac{\bar{n}(\omega)}{\omega+\omega_{qj}}\frac{J^j(\omega)}{2\pi}\right\},\\
\Gamma_{\uparrow x}^{{qj}}& =&J^j(\omega_{qj}) \bar{n}(\omega_{qj})- i \cdot 2 {\cal P}\int_0^{\infty}\di \omega \left\{\frac{\bar{n}(\omega)}{\omega-\omega_{qj}}\frac{J^j(\omega)}{2\pi}-\frac{\bar{n}(\omega)+1}{\omega+\omega_{qj}}\frac{J^j(\omega)}{2\pi}\right\},\\
\Gamma_{\downarrow X}^{{qj}}& =&J_{X}^j(\omega_{qj}) [\bar{n}(\omega_{qj})+1]+ i \cdot 2 {\cal P}\int_0^{\infty}\di \omega \left\{\frac{\bar{n}(\omega)+1}{\omega-\omega_{qj}}\frac{J_X^{j}(\omega)}{2\pi}-\frac{\bar{n}(\omega)}{\omega+\omega_{qj}}\frac{[J_X^j(\omega)]^*}{2\pi}\right\},\\
\Gamma_{\uparrow X}^{{qj}}& =&[J_{X}^j(\omega_{qj})]^* \bar{n}(\omega_{qj})- i \cdot 2 {\cal P}\int_0^{\infty}\di \omega \left\{\frac{\bar{n}(\omega)}{\omega-\omega_{qj}}\frac{[J_X^j(\omega)]^*}{2\pi}-\frac{\bar{n}(\omega)+1}{\omega+\omega_{qj}}\frac{J_X^j(\omega)}{2\pi}\right\}.
\eea
This is the most general form of the second order master equation for two systems coupled to the same reservoir. (The master equation can be put in Lindblad form but we will only do this for the simpler case of two excitons). 

The above master equation is complete but it also contains terms that contribute very little to the dynamics and a simplified master equation can be obtained by `secularizing' the master equation as outlined in \cite{eastham2013lindblad}. Secularization is an approximation akin to a rotating wave approximation (RWA), which allows us to eliminate the incoherent terms in the master equation which oscillate a lot faster compared to the system timescales. Equivalently, we can say that secularization allows us to eliminate the incoherent terms in the master equation which do not conserve energy and therefore are forbidden to second order.
 We expect that, similarly to a RWA, this approximation holds as long as the strength of the terms ignored is small compared to their oscillation frequency. 

\subsection{Two un-coupled driven excitons interacting with a common phonon bath}
In the case of a pair of un-coupled excitons coupled to the same phonon bath and to separate photon baths as in Eqs.~\ref{StartingHamiltonian} and \ref{Hamiltonian} we have the reservoir-system coupling operator $(S_j+S_j^\dagger)=\tilde{c}_j^\dagger \tilde{c}_j$ which yields $S_j=\frac{\sin\theta}{2}\sigma_-^j+\frac{1}{4}(\sigma_z^j\cos\theta+I)$. We also have $g_{\bm{k}}=t_{\bm{k}} e^{i \bm{k} \cdot \bm{d}}$\cite{kok2010introduction}.

The starting Hamiltonian is:
\bea
H&=&\frac{W_A}{2}\sigma_z^A +\frac{ W_B}{2}\sigma_z^B + \sum_{\bm{k}} \omega_{\bm{k}} b_{\bm{k}}^\dagger b_{\bm{k}}+H_\gamma\\
& &+\sum_{\bm{k}} (b_{\bm{k}}+b_{\bm{k}}^\dagger)\left[t_{\bm{k}}\left(\frac{\sin\theta}{2}\sigma_-^A+\frac{1}{4}(\sigma_z^A\cos\theta+I)\right)+h.c.\right]\nonumber\\
& &+\sum_{\bm{k}} (b_{\bm{k}}+b_{\bm{k}}^\dagger)\left[t_{\bm{k}} e^{i \bm{k}\cdot \bm{d}}\left(\frac{\sin\theta}{2}\sigma_-^B+\frac{1}{4}(\sigma_z^B\cos\theta+I)\right)+h.c.\right],\\
\eea
where we define $H_{\gamma}=\sum_{\bm{q}} \Theta_{\bm{q}} a^\dagger_{\bm{q}}a_{\bm{q}}+\sum_j (e^{i\omega_l^j t}\tilde{c}^\dagger_j +e^{-i\omega_l^j t }\tilde{c}_j)  \sum_{\bm{q}} f^j_{\bm{q}} (a^\dagger_{\bm{q}} +a_{\bm{q}})$; this is the Hamiltonian resulting from the coupling of the QDs to the radiation field. Since the two QDs probe the radiation field at different frequencies the photon bath does not mediate any interaction between the two QDs and therefore we can safely treat the photon bath as effectively two photon baths interacting separately and individually with each QD. We assume that the reader is familiar with the usual procedure of tracing out the photon operators $a_{\bm{q}}$ to obtain the contribution to the second-order Born-Markov master equation and therefore directly insert the resulting dissipators in Equation \ref{GeneralMaster} below.

We define the following spectral density functions: 
\bea
J_p(\omega) &=&\sum_{\bm{k}} \abs{t_{\bm{k}}}^2 \delta(\omega-\omega_{\bm{k}}),\\
J_\gamma^j(\omega)&=&\sum_{\bm{q}} \abs{f_{\bm{q}}}^2 \delta (\omega-\omega_{\bm{q}}),\\
J_X(\omega) &\equiv&\sum_{\bm{k}} \abs{t_{\bm{k}}}^2 e^{i\bm{k}\cdot \bm{d}} \delta(\omega-\omega_{\bm{k}})=F_{\mathds D}\left(\frac{\omega d}{c_s}\right) J_p(\omega),
\eea
where $c_s$ is the speed of sound in GaAs and $F_{\mathds D}(x)$ is a function that depends on the dimensionality of the phonons such that $F_{\mathds D}(x)=\mathrm{sinc}(x)$ in 3D,  $F_{\mathds D}(x) = J_0(x)$ ($J_0$ is the Bessel function of the first kind) in 2D and $F_{\mathds D}(x) = \exp(i x)$ in 1D. 

Using the above notation we obtain the following secularized master equation in Lindblad form describing the interaction of two excitons with a common phonon bath but separate photon baths:
\bea\label{GeneralMaster}
\dot{\rho}&=&-i\left[H,\rho\right] \\
& &+\sum_j \left\{ \left[\Gamma_\downarrow^j -\abs{ \Re \left(\Gamma_\downarrow^X\right)}-\abs{\Im \left(\delta \Gamma_\downarrow^X\right)}  \right] {\cal D} (\sigma_-^j, \rho)+\left[\Gamma_\uparrow^j -\abs{ \Re \left(\Gamma_\uparrow^X\right)}-\abs{\Im \left(\delta \Gamma_\uparrow^X\right)}\right] {\cal D} (\sigma_+^j, \rho)\right\} \nonumber\\
& &\nonumber+ \abs{\Re\left( \Gamma_\downarrow^X  \right)} {\cal D} (\sigma_-^A\pm \sigma_-^B, \rho)+\abs{\Re\left( \Gamma_\uparrow^X  \right)} {\cal D} (\sigma_+^A\pm \sigma_+^B, \rho)\\
& &\nonumber +\abs{\Im\left( \delta \Gamma_\downarrow^X  \right)} {\cal D} ( i \sigma_-^A\pm \sigma_-^B, \rho)+\abs{\Im\left( \delta \Gamma_\uparrow^X  \right)} {\cal D} (i \sigma_+^A\pm \sigma_+^B,\rho),
\eea
where in the last four terms the sign $\pm$ between the two operators forming the dissipators is the sign of their corresponding rates and where: 
\bea
 H&=& \frac{W_A+\Delta_A}{2} \sigma_z^A  + \frac{W_B+\Delta_B}{2} \sigma_z^B +\frac{\Delta_X}{2}\sigma_z^A\sigma_z^B \\
& &+\frac{\Im\left( \Gamma_\downarrow^X+ \Gamma_\uparrow^X \right)}{2} (\sigma_+^A\sigma_-^B+\sigma_-^A\sigma_+^B)+i \frac{\Re\left(\delta \Gamma_\downarrow^X-\delta \Gamma_\uparrow^X \right)}{2} (\sigma_+^A\sigma_-^B - \sigma_-^A\sigma_+^B)\nonumber\\
\Delta_j&=&\frac{\sin^2\theta_j}{2} {\cal P} \int_0^\infty \di \omega \frac{(2\bar{n}(\omega)+1)W_j}{\omega^2-W_j^2}\frac{J_p(\omega)}{2\pi} + \cos(\theta_j){\cal P}\int_0^\infty \di \omega \frac{1}{\omega} \frac{J_p(\omega) + \Re\{J_X(\omega)\}}{2\pi}\nonumber\\
\Delta_X&=&\cos\theta_A\cos\theta_B\int_0^\infty \frac{1}{\omega} \frac{\Re\{J_X(\omega)\}}{2\pi}\nonumber\\
\Gamma_\downarrow^j &=&\cos^4\left(\frac{\theta_j}{2}\right)J_\gamma^j (\omega_l^j) +   \frac{\sin^2(\theta_j)}{4}J_p (W_j) [\bar{n}(W_j)+1]\nonumber,\\
\Gamma_\uparrow^j &=&\sin^4\left(\frac{\theta_j}{2}\right)J_\gamma^j(\omega_l^j)+\frac{\sin^2(\theta_j)}{4}J_p (W_j) \bar{n}(W_j)\nonumber,\\
\Gamma_\phi^j&=&\frac{\sin^2\theta_j}{4} J_\gamma^j(\omega_l^j)\nonumber,\\
\Gamma_\downarrow^X&=&\frac{\Gamma_{\downarrow X}^A+\Gamma_{\downarrow X}^B}{2},\quad\Gamma_\uparrow^X=\frac{\Gamma_{\uparrow X}^A+\Gamma_{\uparrow X}^B}{2},\quad \delta \Gamma_{\downarrow}^X=\frac{\Gamma_{\downarrow X}^A-\Gamma_{\downarrow X}^B}{2},\quad \delta \Gamma_\uparrow^X=\frac{\Gamma_{\uparrow X}^A-\Gamma_{\uparrow X}^B}{2}\nonumber,
\eea
and where we have defined: 
\bea
\Gamma_{\downarrow X}^{j}& =&\frac{\sin\theta_A\sin\theta_B}{4}\left\{J_{X}^j(W_j) [\bar{n}(W_j)+1]+ i \cdot 2 {\cal P}\int_0^{\infty}\di \omega \left[\frac{\bar{n}(\omega)+1}{\omega-W_j}\frac{J_X^j(\omega)}{2\pi}-\frac{\bar{n}(\omega)}{\omega+W_j}\frac{[J_X^j(\omega)]^*}{2\pi}\right]\right\}\nonumber.\\
\Gamma_{\uparrow X}^{j}& =&\frac{\sin\theta_A\sin\theta_B}{4}\left\{[J_{X}^j(W_j)]^* \bar{n}(W_j)- i \cdot 2 {\cal P}\int_0^{\infty}\di \omega \left[\frac{\bar{n}(\omega)}{\omega-W_j}\frac{[J_X^j(\omega)]^*}{2\pi}-\frac{\bar{n}(\omega)+1}{\omega+W_j}\frac{J_X^j(\omega)}{2\pi}\right]\right\}\nonumber.
\eea

Since the most pronounced effects are at resonance when $W_A\approx W_B$ in analytical calculations we ignore the terms proportional to $\delta \Gamma_{\downarrow,\uparrow}^X$ in \ref{GeneralMaster}, although we use the most general master equation in numerical simulations.

\section*{References}
\bibliographystyle{unsrt}

\end{document}